\documentstyle[aps,prl,epsf,twocolumn,floats,amssymb,amsfonts]{revtex}

\begin{document}
\draft
\title{
Localization and delocalization in dirty superconducting 
wires}

\author{P.\ W.\ Brouwer$^{a}$, A.\ Furusaki$^{b}$, I.\ A.\ Gruzberg$^{c}$, 
and C.\ Mudry$^{d}$}
\address{$^{a}$Laboratory of Atomic and Solid State Physics,
Cornell University, Ithaca, NY 14853-2501\\
$^{b}$Yukawa Institute for Theoretical Physics, Kyoto University,
Kyoto 606-8502, Japan\\
$^{c}$Institute for Theoretical Physics, University of California, 
Santa Barbara, CA 93106-4030\\
$^{d}$Paul Scherrer Institut, CH-5232, Villigen PSI, 
Switzerland\\
{\rm February 1, 2000}
\medskip \\ \parbox{14cm}{\rm 
We present Fokker-Planck equations that describe transport of heat
and spin in dirty unconventional superconducting quantum wires.
Four symmetry classes are distinguished, depending on the presence
or absence of time-reversal and spin rotation invariance. In the absence 
of spin-rotation symmetry, heat transport is anomalous in that the 
mean conductance decays like $1/\sqrt{L}$ instead of exponentially 
fast for large enough length $L$ of the wire. 
The Fokker-Planck equations in the presence of time-reversal symmetry
are solved exactly and the mean conductance for quasiparticle
transport is calculated for the crossover from the diffusive to the
localized regime.
}}

\maketitle 


\narrowtext

The discovery of the $d$-wave nature of the order parameter in high
$T_c$ materials has renewed interest in unconventional superconductors
with low energy quasiparticles near the Fermi energy $\varepsilon_F$.
An important question is how disorder affects the quasiparticle
dynamics and the corresponding low-temperature properties of the
superconductor. In Ref.\ \onlinecite{Gorkov} it is predicted that, on
energy scales $|\varepsilon- \varepsilon_F|$ less than the inverse mean
free time and on length scales beyond the mean free path, weak
impurity scattering leads to a finite density of states (DoS)
and to a diffusive dynamics of quasiparticles. In normal metals, it
has been known for a long time that quantum interference imposes
corrections to this picture, in the form of weak localization, and
eventually, for dimensions $\le 2$, exponential (Anderson)
localization.  The analogous question for low energy quasiparticles in
unconventional superconductors has been considered only recently
\cite{HigherD,Bundschuh,Senthil,spinQH,classD,Bocquet}.

The crucial distinction between quasiparticles in a superconductor and
in a normal metal, is that the former are described by a Hamiltonian
of Bogoliubov-de Gennes (BdG) type. Such a Hamiltonian has an
additional particle-hole grading, accompanied by a discrete
particle-hole symmetry, which is absent in the Hamiltonian for
(electron-like) quasiparticles in a normal metal. Symmetry plays a
crucial role in the problem of Anderson localization. A classification
of the symmetry classes for BdG Hamiltonians, depending on the
presence or absence of time-reversal (TR) and spin-rotation (SR) symmetry,
has been given by Altland and Zirnbauer \cite{AltlandZirnbauer}. The
four possibilities are denoted C, CI, D, and DIII, see table
\ref{tab:1}. Ref.\ \onlinecite{AltlandZirnbauer} addressed the
``zero-dimensional'' 
(0D)
case of chaotic quantum dots with
superconducting leads. The higher dimensional realizations of the BdG
symmetry classes, relevant for the question of localization, were
studied in Refs.\ 
\onlinecite{HigherD,Bundschuh,Senthil,spinQH,classD,Bocquet}, mainly
by field-theoretical methods involving construction and analysis of
non-linear sigma models with appropriate symmetries.

In this Letter we study localization in the BdG symmetry classes for
the geometry of a quantum wire, i.e., in quasi-one-dimension (quasi-1D). 
For this
purpose, we use the Fokker-Planck (FP) approach
\cite{DorokhovMPK,Transport}, which is complementary to the non-linear
sigma model of Refs.\
\onlinecite{HigherD,Bundschuh,Senthil,classD,Bocquet}. Using
the
classification scheme of Ref.\ \onlinecite{AltlandZirnbauer}, we
obtain four FP equations that control quasiparticle
transport at the Fermi level in a dirty superconducting wire. Our
findings are remarkable: While for classes C and CI the mean and
typical values of the quasiparticle conductance $g$ decay
exponentially with the length $L$ of the wire for large $L$, the
situation in classes D and DIII is quite
different. There the mean $\langle g \rangle$ decays only
algebraically to zero for large $L$ and $\ln g$ is not self-averaging,
indicating a very broad distribution of the conductance and the
absence of the exponential localization of the quasiparticle states
at $\varepsilon_F$. (The absence of exponential localization for 
class D has been announced independently in Ref.\ \onlinecite{Bocquet}.)

It should be stressed that the BdG Hamiltonians do not conserve
charge.  Instead, the conserved densities are those of the energy (in
all four classes) and spin (when the SR symmetry is present). Thus,
the transport properties (the conductance $g$) studied in this
Letter refer to transport of heat and spin.

We now proceed with a detailed statement of our results and their
derivation. The model that we consider is that of a disordered quantum
wire, with a Hamiltonian of the BdG form. We distinguish gradings
corresponding to spin up/down, particle/hole, left/right
movers. Denoting these with Pauli matrices $\sigma$, $\gamma$, and
$\tau$, respectively, we write our model Hamiltonian as
\label{eq:our model}
\begin{equation}
{\cal H} = {\cal K} + {\cal V}, \qquad 
{\cal K} = i v_F \partial_x
  \sigma_0 \otimes \gamma_0 \otimes \tau_3 \otimes \openone_{N},
  \label{eq:cal H}
\end{equation}
where $\sigma_0$ is the $2 \times 2$ unit matrix in the spin grading
etc. The kinetic energy ${\cal K}$ describes the propagation of right
and left moving quasiparticles in $N$ channels at the Fermi level. The
``potential'' ${\cal V}(x)$ is an $8N \times 8N$ matrix that accounts
both for the presence of disorder and of superconducting
correlations. In particle/hole ($\gamma$) grading it reads
\begin{equation}
  {\cal V} =
\pmatrix{
v&\Delta\cr
-\Delta^*&-v^{\rm T}\cr
},
\label{eq:def V}
\end{equation}
where $v$ ($\Delta$) is a hermitian (antisymmetric) $4N \times 4N$
matrix, representing the impurity potential (superconducting order
parameter).  The form (\ref{eq:def V}) of the potential ${\cal V}$
ensures that the Hamiltonian ${\cal H}$ obeys particle-hole symmetry,
${\cal H} = - \gamma_1 {\cal H}^{\rm T} \gamma_1$ \cite{AltlandZirnbauer}.
In addition, ${\cal H}$ (and hence ${\cal V})$ may obey TR
invariance ${\cal H} = {\cal T}{\cal H}^* {\cal T}^{-1}$, with ${\cal
T}=i\tau_1\otimes\sigma_2$, and/or SR invariance
${\cal H} = -\gamma_2 {\cal H}^{\rm T} \gamma_2$.  

\begin{table}
\begin{tabular}{l|c|c|c|c|c|c|c}
Class    & {SR} & {TR} & ${\cal L}$             & 
  ${\cal G}$                      & $m_0$ & $m_l$ & $d$ \\ \hline
{ C}     & Yes  & No   & ${\rm Sp}(N,N)$        &
  ${\rm Sp}(N)\times {\rm Sp}(N)$ &  4    &  3    &  4  \\
{ CI}    & Yes  & Yes  & ${\rm Sp}(N,{\Bbb C})$  &
  ${\rm Sp}(N)                  $ &  2    &  2    &  4  \\
{ D}     & No   & No   & ${\rm O}(4N,4N)$       & 
  ${\rm O}(4N)\times {\rm O}(4N)$ &  1    &  0    &  1  \\
{ DIII}  & No   & Yes  & ${\rm O}(4N,{\Bbb C})$  &
  ${\rm O}(4N)$                   &  2    &  0    &  2  \\
\end{tabular}\hfill\\
\caption{\label{tab:1}
Definition of the symmetry classes for a dirty superconducting quantum
wire, with respect to the presence of spin-rotation (SR) and
time-reversal (TR) symmetry.  The table lists the Lie group ${\cal L}$
of the transfer matrix ${\cal M}$, the factor group ${\cal G}$ of
angular degrees of freedom of ${\cal M}$, the multiplicities of the
roots of the symmetric space ${\cal L}/{\cal G}$, and the degeneracies
of the radial coordinates $x_j$ of the transfer matrix ${\cal M}$.}
\end{table}

Spatial fluctuations of the order parameter $\Delta$ and the potential
$v$ are taken into account by assuming that ${\cal V}$ is a Gaussian
random variable with vanishing mean, i.e.\ that its probability
functional $P[{\cal V}]$ is of the form
\begin{equation}
  P[{\cal V}] \propto \exp\left[-{\gamma \ell \over 4c}\int_0^L dx\,
    {\rm tr}\,{\cal V}^2 (x)\right],
  \label{eq:prob dist}
\end{equation}
where $\ell$ is the mean free path, $\gamma$ is a numerical
constant to be defined below, and $c=1$ ($2$) for class C/D (CI/DIII).  
The transport properties of the
Hamiltonian ${\cal H}$ describe the transport of spin and heat by
quasiparticles in a disordered superconducting quantum wire.

Before we continue with the analysis of our model (\ref{eq:cal
H}--\ref{eq:prob dist}), some remarks about its validity and relevance
are in place. One key property of the model is that, apart from 
corrections at very low energies due to quasiparticle localization
\cite{Senthil} or the appearance of a critical state, the DoS of the
Hamiltonian ${\cal H}$ near the Fermi level $\varepsilon_F$ is
nonzero and finite.
This is related to the fact that the statistical average of
the order parameter $\Delta$ is zero in our model, cf.\ Eq.\
(\ref{eq:prob dist}).  For a dirty superconductor, such behavior is
plausible if the order parameter is unconventional, as in
$d$-wave superconductors, or when it breaks TR
symmetry, as is believed to be the case for, e.g., the
ruthenates \cite{Ruthenates}, vortex lines in a (conventional)
superconductor \cite{Bundschuh}, 
or a normal metal wire with magnetic impurities that is
weakly connected to a superconducting substrate.
In all these cases the disorder leads to the existence of low-energy 
quasiparticle states \cite{Gorkov,Senthil}. 
The Hamiltonian (\ref{eq:cal H}) 
then describes diffusion and localization of these
``disorder-facilitated'' quasiparticles. 
An altogether different scenario is that of a wire made out of an
unconventional superconductor with very weak disorder. If boundary
conditions are suitably chosen, one or several propagating 
modes can exist at $\varepsilon_F$, whose localization properties
are described by Eq.\ (\ref{eq:cal H}). 
In any case, one should view ${\cal H}$ as an effective or coarse
grained Hamiltonian, whose validity is restricted to length scales
beyond the microscopic mean free path $\ell$. It is universal in the
sense that its form is determined solely by the symmetry, and the
distribution (\ref{eq:prob dist}) provides for the existence of the
diffusive regime with a finite DoS at the proper energy scale.  (Note
that the restrictions to the validity of our model are not different
from those of related field theoretic descriptions appearing in the
literature \cite{Bundschuh,Senthil,classD,Bocquet}.)

We describe transport properties of the model (\ref{eq:cal
H}--\ref{eq:prob dist}) through its $8N \times 8N$ transfer matrix
${\cal M}$ that encodes the $x$-dependence of an $8N$-component
quasiparticle wavefunction $\psi$ satisfying the Schr\"odinger
equation ${\cal H} \psi = \varepsilon \psi$ at $\varepsilon=0$,
$
  \psi(x+L) = {\cal M}(x+L,x) \psi(x).
$
Formally, ${\cal M}$ is related to the Hamiltonian (\ref{eq:cal H}) as
\begin{equation}
  {\cal M}(x+L;x)=
  {\rm T}_{y}
  \exp
  \left[
  i\int_x^{x+L} dy\, \tau_3\, {\cal V}(y) \right],
\label{eq:def cal M}
\end{equation}
where ${\rm T}_{y}$ denotes the path ordering operator for the
$y$-integration along the wire. {}From Eq.\ (\ref{eq:def cal M}) one
finds that flux conservation (i.e., Hermiticity of ${\cal H}$) and 
particle-hole symmetry imply that ${\cal M}^{\dagger} \tau_3 {\cal M} =
\tau_3$ and $\gamma_1 {\cal M} \gamma_1 = {\cal M}^*$,
respectively. Further, TR invariance requires $ {\cal T} {\cal M}
{\cal T}^{-1} = {\cal M}^*$, while SR invariance is obeyed if
$\gamma_2 {\cal M} \gamma_2 = {\cal M}^*$.  The transfer matrix ${\cal
M}$ obeys the multiplicative rule ${\cal M}(z,x) = {\cal M}(z,y) {\cal
M}(y,x)$ for $x < y < z$ and hence is an element of a certain Lie
group ${\cal L}$.  The appropriate Lie groups for the four symmetry
classes are listed in Table \ref{tab:1}. We note that the actual
transfer matrix group is an $8N$-dimensional representation of the Lie
group ${\cal L}$, where ${\cal L}$ also allows a lower dimensional
(irreducible) representation for the classes C, CI, and DIII
\cite{LieGroupExample}.  Elements of ${\cal L}$ are conveniently
parameterized in terms of their polar decomposition, which, in an
irreducible representation, takes the form
\begin{eqnarray*}
&
\pmatrix{V_1&0\cr0&\!\! V_2\cr}
\pmatrix{\cosh X&\sinh X\cr\sinh X&\cosh X\cr}
\pmatrix{V_3&0\cr0&\!\! V_4\cr} \ \ & (\mbox{C, D}),
\nonumber \\
&
V_1
\pmatrix{\cosh X&i\sinh X\cr-i\sinh X&\cosh X\cr}
V_2 \ \ & (\mbox{CI, DIII}).
\end{eqnarray*} 
Here $V_i \in {\rm O}(4N)$ [Sp$(N)$] for classes D/DIII [C/CI], for
all $i=1,2,3,4$, and $X$ is a diagonal matrix with positive
entries $x_j$. (By Kramers' degeneracy, the elements
of $X$ occur in pairs in class C.) The $x_j$ serve as radial
coordinates on the Lie Group ${\cal L}$. One verifies that the
eigenvalues of the true $8N \times 8N$ transfer matrix ${\cal M} {\cal
M}^{\dagger}$ occur in $d$-fold degenerate inverse pairs $\exp(\pm 2
x_j)$, where the degeneracy $d$ is listed in Table \ref{tab:1}. Hence
the number of independent $x_j$'s is $4N/d$. Finally, we note that the
$x_j$ are related to the conductance $g$ through \cite{Transport}
\begin{equation}
  g = d \sum_{j=1}^{4N/d} \cosh^{-2} x_j.
\end{equation}

Our aim is to find the probability distribution of the $x_j$ for a
transfer matrix corresponding to the model 
(\ref{eq:cal H}--\ref{eq:prob dist}).
Increasing the length $L$ of the wire by a small increment $\delta L$
amounts to multiplication of its transfer matrix ${\cal M}(L) = {\cal
M}(x+L,x)$ by a transfer matrix ${\cal M}' = {\cal
M}(x+L+\delta L,x+L)$.  Since ${{\cal M}}'$ is close to the unit
matrix, random, and statistically independent from ${\cal M}(L)$, we
find that as a function of $L$, ${\cal M}(L)$ performs a random
trajectory on its Lie group ${\cal L}$. Actually, we do not need to
know the full trajectory on ${\cal L}$ if we are only interested in
the conductance $g$. It is sufficient to know the
trajectory of the radial coordinates $x_j$ of ${\cal M}(L)$ after
dividing out a maximal compact subgroup ${\cal G}$ of ${\cal L}$
corresponding to the angular degrees freedom of ${\cal M}$ that leave
the product ${\cal M} {\cal M}^{\dagger}$ invariant, or, in other
words, to know the trajectory of the $x_j$ in the symmetric space
${\cal L}/{\cal G}$ \cite{Caselle}.  The subgroups ${\cal G}$ are
listed in Table \ref{tab:1}.
Starting from the microscopic model (\ref{eq:cal H}-\ref{eq:prob
dist}), one can show that the trajectory obeyed by the $x_j$ is a
Brownian motion on the coset space ${\cal L}/{\cal G}$ described by
the joint probability distribution $P(x_1,\ldots,x_{4N/d};L)$. The
$L$-evolution of $P$ is described by a 
FP equation, which
follows either from a direct calculation starting from Eq.\
(\ref{eq:cal H}), or from the general theory of symmetric spaces
\cite{Caselle}. In both cases we find
\begin{eqnarray}
{\partial P\over\partial L} &=&
{1\over2 \gamma \ell}
\sum_{j=1}^{4N/d}
{\partial\over\partial x_j}
\left[
J
\left(
{\partial\over\partial x_j} J^{-1}P
\right)
\right],
\label{eq:FPbdg}\\
J &= &
\prod_{j=1}^{4N/d}
|\sinh 2x_j|^{m_l}
\prod_{k>j}^{4N/d} \prod_{\pm}
|
\sinh(x_j\pm x_k)
|^{m_o}, \nonumber
\end{eqnarray}
where the numbers $m_{l}$ and $m_{o}$ are the long and ordinary
root multiplicities in the symmetric spaces ${\cal L}/{\cal G}$,
see Table \ref{tab:1}, and $\gamma = (4N m_o/d) + 1 - m_o + m_l$.
The FP equation (\ref{eq:FPbdg}) is supplemented with
the boundary condition ${\partial P/\partial x_j} = 
(P/J) {\partial J / \partial x_j}$ at $x_j = 0$.
The initial condition ${\cal M} = 1$ for $L=0$ corresponds to
$P(x_1,\ldots,x_{4N/d};0) = \prod_{j} \delta(x_j)$.

The FP equation (\ref{eq:FPbdg}) is the fundamental
equation that governs quasiparticle transport and localization in
quantum wires of the symmetry classes C, CI, D, and DIII.  

In the localized regime $L \gg N\ell$, typically all $x_j$ and their
spacings are much bigger than unity, and the conductance is governed
by the smallest coordinate $x_1$. That coordinate has a Gaussian
distribution, with mean $m_l L/\gamma \ell$ and variance $L/\gamma
\ell$. For classes C and CI this implies that 
$g$ is exponentially small, with
\begin{equation}
  \langle \ln g \rangle = - {2 m_l L \over \gamma \ell}, \qquad
  \mbox{var}\, \ln g = {4 L \over \gamma \ell},\ \ 
\end{equation}
with $m_l = 2$ for class CI and $m_l = 3$ for class C. Exponential
localization for class C in quasi-1D was previously obtained by 
Bundschuh {\em et al.} \cite{Bundschuh}, using the non-linear sigma
model.
For class D and DIII however, $m_l = 0$, so that there is no exponential 
localization. Instead, $g$ has a very broad distribution (broader than  
log-normal), with an algebraic decay of the mean and the variance
and an $L^{1/2}$-dependence of $\ln g$,
\begin{eqnarray}
  && \langle g \rangle  = d \sqrt{2 \gamma \ell \over \pi L}, \qquad
  \mbox{var}\,  g = {2 d \over 3} \langle g \rangle,
  \nonumber \\
\label{eq:localizedD}
  && \langle \ln g \rangle = -4 \sqrt{L \over 2 \pi \gamma \ell}, \qquad
  \mbox{var}\, \ln g = {4 (\pi - 2) L \over \pi \gamma \ell}.
\end{eqnarray}
Hence in classes D and DIII, quasiparticle states are not localized at
the Fermi level. Since they are neither truly extended (typically $g
\ll 1$ in class D and DIII), we label them critical, following
terminology from the case of quantum wires with off-diagonal disorder,
where similar behavior is found at the center of
the band \cite{Dyson53,BMSA}.

The effect of disorder is much less pronounced in the 
diffusive regime $\ell\ll L\ll N\ell$. Here, the conductance has only small 
fluctuations around its mean. Following the method of 
moments \cite{Transport}, we find $\langle g \rangle$ from the FP 
equation by construction of
evolution equations for the moments of $g_a=d \sum_j\cosh^{-2a} x_j$,
$a=1,2,\ldots$,
\label{eq:dif D DIII}
\begin{eqnarray}
{\gamma \ell \over a}{\partial
\left\langle
g_a
\right\rangle
\over
\partial L
}
&=&
{m_o \over d}\sum_{n=1}^{a-1} \langle g_{a-n} g_{n} \rangle
  - {m_o \over d} \sum_{n=1}^{a} \langle g_{a-n+1} g_{n} \rangle
  \nonumber \\ && \mbox{}
  + (a m_o - 2a-1 + m_l) \langle g_{a+1} \rangle
  \nonumber \\ && \mbox{}
  + (2a- a m_o + m_o - 2 m_l) \langle g_{a  } \rangle.
\end{eqnarray}
In the diffusive regime one may replace the average of a product by
the product of the averages, and hence one finds for $\ell \ll L \ll
N \ell$
\begin{eqnarray} \label{eq:WL}
  \langle g \rangle = {4N\ell \over L + \ell} + 
  {d (m_o - 2 m_l) \over 3 m_o} + {\cal O}(\ell/L,L/N \ell).
\end{eqnarray} 
The first term in Eq.\ (\ref{eq:WL}) is the Drude conductance, while
the second term is the first quantum interference correction to the
average conductance. For classes C, CI, D, and DIII it takes the
values $-2/3$, $-4/3$, $1/3$, and $2/3$, respectively. The weak
localization correction for class C was obtained earlier in Ref.\
\onlinecite{Bundschuh}. For the classes D and DIII, the correction is
positive, i.e.,\ quantum interference enhances the conductance
relative to the classical Drude-like leading behavior (see also 
Ref.\ \onlinecite{Bundschuh}). This
is similar to the phenomenon of anti-localization in the
standard symplectic symmetry class, though, as pointed out by Bocquet
et al.\cite{Bocquet}, here it is a precursor of the breakdown of
exponential localization, while in the standard symplectic class
localization takes over in higher order quantum corrections.

In the presence of TR symmetry, Eq.~(\ref{eq:FPbdg}) is soluble. This
is in contrast to the case of the FP equations for the standard and
chiral symmetry classes, where only the case of broken TR symmetry was
exactly solvable \cite{Transport,BMSA}.  The first step
\cite{Transport} is a map of Eq.~(\ref{eq:FPbdg}) onto a Schr\"
odinger equation in imaginary time for the wave function
$\Psi(\{x_j\};s) = \exp\left[-{1\over2}\ln
J(\{x_j\})\right]P(\{x_j\};s)$ for $4N/d$ fermions in one dimension
with coordinates on the half-line $x>0$. They interact through a
two-body potential proportional to $m_o-2$ in the presence of a
one-body potential proportional to $(m_l-2)m_l$.  Hence, for classes
DIII and CI these fermions are free and they only differ by the
boundary condition obeyed by their wave functions $\Psi$ at the
origin.  We thus find the solutions
\begin{eqnarray}
P & \propto &
  \prod_{j} (x_j \sinh 2 x_j)^{m_l/2} e^{- \gamma x^2_j \ell/2L}
  \nonumber \\ && \mbox{} \times
\prod_{j < k} 
\left(x_k^2 - x_j^2 \right) \left(\sinh^2 x_k - \sinh^2 x_j 
\right).
\end{eqnarray}
Using the method of bi-orthogonal functions\cite{MuttalibFrahm} 
it is then possible to calculate the average conductance $\langle g \rangle$
for all $N$ and $L$. Here we report the result for the limit of large
$N$, leaving the results for finite $N$ for a future 
publication,
\begin{eqnarray}
\langle g\rangle &=&
{1 \over s}
-
{4\over3}
+ 4
\sum_{n=1}^{\infty} e^{-\pi^2 n^2/4s} \left( {1 \over s}  + {2 \over \pi^2 
n^2}\right) \ \ \mbox{CI}, \nonumber\\
\langle g\rangle &=&
{1 \over s}
+
{2\over3}
-
4 \sum_{n=1}^{\infty} e^{-\pi^2n^2/2s} {1 \over \pi^2 n^2}  \ \ \ \ \ \
 \ \ \ \ \ \ \ \mbox{DIII},
\label{eq:exact}
\end{eqnarray}
where $s = L/(4 N \ell)$. Note the agreement with Eq.\ (\ref{eq:WL})
in the diffusive regime $s \ll 1$. In the
localized regime $s \gg 1$, Eq.\ (\ref{eq:exact}) may be resummed, and
the asymptotic result (\ref{eq:localizedD}) is reproduced for class DIII,
while for class CI one finds $\langle g \rangle = 8(\pi s)^{-1/2} \exp(-4 s)$.
Quite remarkably, the exact results (\ref{eq:exact}) for $\langle g \rangle$ 
in classes CI and DIII are related to the average conductance $\langle g 
\rangle_{\rm ch}$ in the chiral unitary ensemble for odd channel 
number\cite{BMSA} as $\langle g (s/2) \rangle_{\rm CI} + 2 \langle g (s) 
\rangle_{\rm DIII} = 4 \langle g (s) \rangle_{\rm ch}$. 

The absence of localization in wires of classes D and DIII may have
important implications for higher dimensions, provided our results can
be extended beyond 1D, and provided they are not restricted to the
regime of weak disorder. With respect to the latter restriction, we
can point to the close formal similarity of the delocalization for the
FP equations of class D/DIII and the 
corresponding 
FP equation for the
chiral symmetry classes with odd $N$, where it is understood that the
absence of localization holds both for weak and strong disorder
\cite{Dyson53,Fisher}. Thus arguing that quasiparticle states at the Fermi
level remain delocalized for arbitrary disorder strength and
dimensionality in the D-classes, our result suggests a possible
resolution of a controversy in the literature surrounding 
2D disordered superconductors of class
D \cite{classD,Bocquet}. While all Refs.\ \onlinecite{classD,Bocquet}
assumed existence of two localized phases, distinguished by the
quantized value of the Hall conductivity $\sigma_{xy}$, and a metallic
phase,
the proposed
global phase diagrams and 
transitions between the phases differ considerably. We suggest that
the solution might simply lie in the absence of
localized phases for classes D and DIII in any dimension $\ge 1$.

In conclusion, we considered quasiparticle transport and localization
in disordered quasi-1D superconducting wires at the Fermi level for the four
Bogoliubov-de Gennes symmetry classes C, CI, D, and DIII. We obtained
and solved the Fokker-Planck equations for the probability of the radial
coordinates of the transfer matrix. While quasiparticle states are
localized in classes C/CI, localization is absent if
spin-rotation symmetry  is broken (classes D/DIII).

We thank A.\ Altland, L. Balents, M.\ P.\ A.\ Fisher, N.\
Read, T.\ Senthil, and M.\ Sigrist for valuable discussions. PWB
gratefully acknowledges that this problem was suggested to him in an
earlier stage by A.\ Altland. Close to completion of this work, we
learned that J.\ T.\ Chalker and coworkers obtained independently
similar results for class D, see also Ref.\ \onlinecite{Bocquet}.
This work was supported by a Grant-in-Aid for Scientific Research from
Japan Society for the Promotion of Science No.\ 11740199 (AF), and by
the NSF under grant No.\ DMR-9528578 (IAG).

\narrowtext


\vspace{-0.6cm}

\begin{references}\vspace{-1.6cm}

\bibitem{Gorkov} L. P. Gorkov and P. A. Kalugin, 
	JETP Lett. {\bf 41}, 253 (1985);
        S. Schmitt-Rink {\em et al.}, 
        Phys. Rev. Lett. {\bf 57}, 2575 (1986);
	P. A. Lee, {\it ibid.} {\bf 71}, 1887 (1993).


\bibitem{HigherD}
	A. Altland {\em et al.}, 
        JETP Lett. {\bf 67}, 22 (1998).

\bibitem{Bundschuh}
	R. Bundschuh {\it et al.}, 
        Nucl.\ Phys.\ B {\bf 532}, 689 (1998); 
	Phys. Rev. B {\bf 59}, 4382 (1999).

\bibitem{Senthil} 
	T. Senthil {\it et al.}, Phys. Rev. Lett. {\bf 81}, 4704 (1998);
	T. Senthil and M. P.\ A. Fisher, Phys. Rev. B {\bf 60}, 6893 (1999). 

\bibitem{spinQH} 
	V. Kagalovsky {\it et al.}, Phys. Rev. Lett. {\bf 82}, 3516 (1999); 
	T. Senthil {\em et al.}, 
        Phys. Rev. B {\bf 60}, 4245 (1999);
	I. A. Gruzberg {\em et al.}, 
        Phys. Rev. Lett. {\bf 82}, 4524 (1999).
        Y.\ Morita and Y.\ Hatsugai, cond-mat/9907001. 
\bibitem{classD}
	T. Senthil and M. P. A. Fisher, cond-mat/9906290;
	N. Read and D. Green, cond-mat/9906453.

\bibitem{Bocquet}
        M.\ Bocquet {\em et al.}, 
        cond-mat/9910480.

\bibitem{AltlandZirnbauer}
        A. Altland and M. R. Zirnbauer, Phys. Rev. Lett. {\bf 76}, 3420 
        (1996); Phys. Rev. B {\bf 55}, 1142 (1997).

\bibitem{DorokhovMPK} 
        O.\  N.\  Dorokhov, 
        JETP Letters {\bf 36}, 318 (1982);
        P.\  A.\  Mello {\em et al.}, 
        Ann.\  Phys.\ (NY) {\bf 181}, 290 (1988).

\bibitem{Transport} 
        C.\ W.\ J.\ Beenakker, Rev.\ Mod.\ Phys.\ {\bf 69}, 731 (1997).


\bibitem{Ruthenates}
        G.\ M.\ Luke {\em et al.}, Nature (London) {\bf 394}, 558 (1998);
        M.\ Sigrist {\em et al.}, Physica C {\bf 317-318}, 134 (1999).


\bibitem{LieGroupExample} 
         For instance, in the presence of
         SR invariance, ${\cal M}$ separates into two identical
         blocks of size $4N \times 4N$. 

\bibitem{Caselle} 
        A. H\"uffmann, J. Phys. A {\bf 23}, 5733 (1990);
        M. Caselle, cond-mat/9610017.

\bibitem{Dyson53} 
        F.\ J.\ Dyson, Phys.\ Rev.\ {\bf 92}, 1331 (1953);
        A.\ D.\ Stone and J.\ D.\ Joannopoulos,
        Phys. Rev. B {\bf 24}, 3592 (1981).

\bibitem{BMSA} 
        P.\ W.\  Brouwer {\it et al.}, 
        Phys.\ Rev.\ Lett.\ {\bf 81}, 862 (1998);
        C.\ Mudry {\em et al.},
        Phys. Rev. B {\bf 59}, 13\,221 (1999).

\bibitem{MuttalibFrahm}
        K.\ A.\ Muttalib, J.\ Phys.\ A {\bf 28}, L159 (1995);
        K.\  Frahm, Phys.\ Rev.\ Lett.\ {\bf 74}, 4706 (1995).

\bibitem{Fisher} 
       D. S. Fisher, Phys. Rev. B {\bf 50}, 3799 (1994); 
       {\bf 51}, 6411 (1995). 

\end{references}
\end{document}